\documentclass[12pt,preprint]{aastex}

\newcommand \chandra{{\it Chandra\/}}  

\newcommand \rosat{{\it ROSAT\/}}
\newcommand \xmm{{\it XMM-Newton\/}}
\newcommand \exosat{{\it EXOSAT\/}} 
\newcommand \swift{{\it Swift\/}}
\newcommand \maxi{{\it MAXI\/}}
\newcommand \rosita{{\it ROSITA\/}}
\newcommand \heao{{\it HEAO-1\/}}
\newcommand \lobster{{\it Lobster-ISS\/}}
\newcommand \msun{\hbox{$\hbox{M}_{\odot}$}}	 
\newcommand \abar{$\bar{A}$}

\slugcomment{Accepted for publication in The Astronomical Journal}
\shorttitle{Large Amplitude X-ray Outbursts}
\shortauthors{DONLEY ET~AL.}

\begin{document}

\title{Large-Amplitude X-ray Outbursts from Galactic Nuclei: A Systematic 
Survey Using \rosat\ Archival Data}

\author{
J. L.~Donley,\altaffilmark{1}$^{,}$\altaffilmark{2} 
W. N.~Brandt,\altaffilmark{1} 
Michael Eracleous,\altaffilmark{1}$^{,}$\altaffilmark{3}
and 
Th. Boller\altaffilmark{4} } 

\altaffiltext{1}{Department of Astronomy \& Astrophysics, 525 Davey
Laboratory, The Pennsylvania State University, University Park, PA 16802}
\altaffiltext{2} {NSF-supported undergraduate research associate}
\altaffiltext{3} {Visiting astronomer, Kitt Peak National Observatory and Cerro Tololo
Inter-American Observatory, which are operated by AURA, Inc., under a
cooperative agreement with the National Science Foundation.}
\altaffiltext{4}{Max-Planck-Institut f\"{u}r extraterrestrische Physik, Postfach 1312,
 85741 Garching, Germany}


\begin{abstract}
      In recent years, luminous X-ray outbursts with variability
      amplitudes as high as $\approx 400$ have been serendipitously
      detected from a small number of active and inactive
      galaxies. These outbursts may result from the tidal disruptions
      of stars by supermassive black holes as well as accretion-disk
      instabilities. In order to place the first reliable constraints
      on the rate of such outbursts in the Universe and test the
      stellar tidal disruption hypothesis, we have performed a
      systematic and complete survey for them by cross-correlating
      \rosat\/ All-Sky Survey (RASS) and pointed PSPC data. We have
      detected five galaxies that were in outburst during the RASS,
      three of which show no signs of nuclear activity; these objects
      had been reported on individually in previous studies.  After
      making reasonable corrections for the complicated selection
      effects, we conclude that the rate of large-amplitude X-ray
      outbursts from inactive galaxies in the local Universe is
      $\approx 9.1 \times 10^{-6}$~galaxy$^{-1}$~yr$^{-1}$. This rate
      is consistent with the predicted rate of stellar tidal
      disruption events in such galaxies. When only the two active
      galaxies are considered, we find a rate for active galaxies of
      $\approx 8.5 \times 10^{-4}$~galaxy$^{-1}$~yr$^{-1}$. In order
      to place tighter constraints on these rates, additional
      outbursts must be detected.

\end{abstract}

\keywords{
galaxies: active ---
galaxies: nuclei ---
X-rays: galaxies}


\section{Introduction}
\label{intro}

The detection of large-amplitude X-ray outbursts originating from
inactive and active galactic nuclei (AGN) has generated considerable
interest. These outbursts have variability amplitudes up to a factor
of $\approx 400$, decay over periods of months to years, often exceed
outburst X-ray luminosities of $10^{43}$--$10^{44}$~erg~s$^{-1}$, and
usually have extremely soft X-ray spectra.  Possible mechanisms
considered to explain these events include (1) the tidal disruptions
of stars by supermassive black holes, (2) accretion-disk
instabilities, and (3) the X-ray afterglows of gamma-ray bursts (e.g.,
Komossa \& Bade 1999 and references therein). Many of the outbursts
detected thus far are best explained by the stellar tidal disruption
scenario.  In addition, at least one of these outbursts appears to be
consistent with the disruption of a brown dwarf or giant planet (Li,
Narayan, \& Menou 2002).  A typical inactive galaxy is expected to
undergo a tidal disruption event as often as every $\approx
10^{4}$--$10^{5}$~yr (e.g., Magorrian \& Tremaine 1999). The resulting
emission from such an event should peak in the extreme ultraviolet or
soft X-ray band, and it should decline over a period of roughly
several months (e.g., Gurzadyan \& Ozernoy 1980; Rees 1990). Of
course, the observable effects of these events remain fairly uncertain
due to the complexity of the disruption and subsequent accretion
processes; an alternate spectral distribution in which significant
emission is radiated in the optical band has been investigated by Loeb
\& Ulmer (1997). Sembay \& West (1993) predicted that at least several
hundred and perhaps as many as several thousand tidal disruption
events should have been detected during the \rosat\ All-Sky Survey if
a substantial fraction of galaxies contain supermassive black holes
(SMBHs) of masses $10^{7}$--$10^{8}$~\msun; investigations of
correlations between central black hole mass and bulge properties
suggest that the latter is true (e.g., Ferrarese \& Merritt 2000;
Gebhardt et al. 2000). Detecting and understanding these events will
aid in determining the importance of the stellar tidal disruption
process in the fueling of SMBHs and may also lead to a better
understanding of accretion-disk instabilities (e.g., Siemiginowska,
Czerny, \& Kostyunin 1996; Burderi, King, \& Szuszkiewicz 1998). X-ray
outbursts could also be related to the creation of double-peaked
emission lines in AGN (e.g., Syer \& Clarke 1992; Eracleous et
al. 1995; Storchi-Bergmann et al. 1997) and nuclear outbursts seen at
other wavelengths (e.g., Cappellari et al. 1999; Renzini et al. 1995).

The first large-amplitude X-ray outburst was detected in the galaxy
E1615+061 using \heao\ and \exosat\ (Piro et al. 1988).\footnote{We
note, however, that a possible detection of a tidal disruption event
was made in 1890 through the visual observation of the galaxy NGC~1068
(Packer 1891; de Vaucouleurs 1991).} All subsequent X-ray detections
have been made using data from \rosat.  \rosat\ was especially
sensitive in the soft ($0.1$--$1.0$~keV) band where the emission from
these outbursts is expected to be strong, and it covered $\approx
20$\% of the sky at least twice. Because constant monitoring of the
X-ray sky or the comparison of at least two observations are the only
ways to detect long-lived outbursts,
\rosat\ provided an excellent means by which large-amplitude X-ray
outbursts could be found and investigated.

Of those galaxies caught undergoing X-ray outbursts, only E$1615+061$,
IC~3599 (Brandt, Pounds, \& Fink 1995; Grupe et al. 1995a), and
WPVS~007 (Grupe et al. 1995b) showed signs of nuclear activity prior
to or after the outburst.  E$1615+061$ is a Seyfert 1, IC~3599 is a
Seyfert 1.9 (Komossa \& Bade 1999), and WPVS~007 is a Narrow-line
Seyfert 1 (NLS1).  NGC~5905 (Bade, Komossa, \& Dahlem 1996),
RX~J$1242.6-1119$ (Komossa \& Greiner 1999), RX~J$1420.4+5334$
(Greiner et al. 2000), and RX~J$1624.9+7554$ (Grupe, Thomas, \&
Leighly 1999) have optical spectra that show no signs of nuclear
activity. Both NGC~5905 and RX~J$1624.9+7554$ are spiral galaxies; the
former is classified as an HII-type.  RX~J$1242.6-1119$ is a pair of
elliptical or early spiral galaxies likely to be interacting, and
RX~J$1420.4+5334$ is also an elliptical or early spiral galaxy. It is
likely that these four galaxies harbor otherwise dormant SMBHs that
became active in the X-ray band only following transient fueling
events.

The X-ray outbursts detected thus far were found either (1)
serendipitously through X-ray and optical follow-up observations of
galaxies or unidentified objects with soft X-ray spectra and other
interesting properties, or (2) serendipitously from \rosat\ fields
pointed at different targets.  In order to place reliable constraints
on the number of such outbursts that occur in the Universe, more {\it
systematic\/} and {\it complete\/} surveys must be performed.  Here we
present the results of such a survey. By cross-correlating
\rosat\ All-Sky Survey (RASS) and pointed observations, we have
identified all \rosat\ sources at high Galactic latitudes that (1)
were in outburst during the six-month RASS and that (2) had count
rates or upper limits a minimum factor of $20$ lower in pointed
observations taken before or after the RASS. We use the results of
this survey to set the first reliable constraints on the frequency of
large-amplitude X-ray outbursts in the Universe.  These constraints
allow comparison with tidal disruption predictions such as those of
Magorrian \& Tremaine (1999), and they also allow assessment of the
ability of \chandra\/, \xmm\/, and future missions to identify
additional outbursts of this type.  We note that Komossa \& Dahlem
(2001) performed a similar survey for the nearby galaxies in the Ho,
Filippenko, \& Sargent (1995) sample, and did not detect any
additional large-amplitude X-ray outbursts.  This work differs from
ours in that X-ray outbursts were looked for only from previously
known, nearby galaxies, whereas our survey looks for variability from
all RASS sources.

Throughout this paper, values of $H_0= 75$~km~s$^{-1}$~Mpc$^{-1}$ and
$q_0 = 0.5$ have been assumed. Galactic column densities have been
taken from either Heiles \& Cleary (1979) or Stark et al. (1992) as
appropriate.


\section{The Survey}

\subsection{Definition of the Survey}

For the purpose of this survey, we define ``large-amplitude X-ray
outbursts'' to be events in galaxies or quasars that cause their count
rates to vary by a minimum factor of 20 in the 0.1--2.4~keV \rosat\
energy band. A minimum variability factor of 20 was chosen because the
X-ray emission of AGN is known to vary by factors up to $\approx
10$--$15$\ through the ``normal'' processes at work in them; we want
to discriminate against ``normal'' AGN variability. In some extreme
cases, moderately obscured AGN (e.g., Seyfert 1.5, 1.8 and 1.9
galaxies) can vary by factors of $\geq 20$ in the soft X-ray band due
to changes in the amount of absorption along the line of sight (e.g.,
the Seyfert 1.5 galaxy NGC~3516; see Guainazzi, Marshall, \& Parmar
2001 and references therein). This behavior is not likely to be
related to the X-ray outbursts that are the focus of this study, and
below we shall discriminate against such objects to the greatest
extent possible using optical classification information, X-ray
spectral information, and the observed variability
characteristics. While this introduces an element of subjectivity into
our study, obscured objects with possible absorption changes are
fortunately not a major source of confusion; only one such object was
found in our survey, and it is discussed in more detail in \S3.1.2.
For the purpose of excluding the most X-ray luminous young supernova
remnants and X-ray binaries from our survey (e.g., Schlegel 1995;
Makishima et al. 2000), we have also defined a minimum outburst X-ray
luminosity threshold of $1 \times 10^{41}$~erg~s$^{-1}$.

In this survey, we focus solely on galaxies and quasars that were
detected in outburst during the six-month RASS and that were either
detected or undetected in their low states in pointed \rosat\
observations preceding or following the RASS.  We have chosen to focus
on this sample because the sensitivity of the typical pointed
observation greatly exceeds that of the typical RASS observation.  The
pointed observations can therefore be used more effectively than the
RASS observations to determine the low count rates and upper limits of
these objects in quiescence.

We chose to search catalogs of the Position Sensitive Proportional
Counter (PSPC) pointed data, and not the data taken with the High
Resolution Imager (HRI), because the observations with the former
cover $\approx 18$\% of the sky while those with the latter cover only
$\approx 2$\% of the sky. Our search of the cataloged PSPC
observations was limited to radii $\leq 50^{\prime}$ from the center
of the PSPC field.  The point spread function of a source whose
central position is located outside of this radius extends beyond the
detector's area, causing counts to be lost. For PSPC observations
centered $\leq 3^{\prime}$ apart, we searched only the field with the
highest exposure time.  While the addition of several PSPC fields is
possible, the catalogs of pointed sources used in our
cross-correlations do not utilize such added fields.  Our sensitivity
was therefore determined by the most sensitive observation in a given
area of sky. The effect of this selection strategy is negligible as
most of the $\approx 700$ overlapping fields removed have exposure
times that are significantly lower than those of the longest exposures
in each of these fields.  Due to the high absorption column density
and large number of confusing sources in the Galactic plane, we do not
consider in our survey the region of the sky from Galactic latitudes
of $-30^{\circ}$ to $+30^{\circ}$.\footnote{For completeness, however,
we have searched for flaring objects in the Galactic Plane and we
summarize our findings in Appendix A.} All large-amplitude X-ray
outbursts to date have been found outside this region of sky. We have
also removed from our survey 138 PSPC fields whose 0.2--2.4~keV
background count rates are greater than $1.4 \times
10^{-3}$~cts~s$^{-1}$~arcmin$^{-2}$; these fields have reduced
sensitivities due to their higher backgrounds, and their inclusion
would introduce significant additional dispersion into the detection
thresholds. With all of the above restrictions taken into account, our
survey has searched $N_{\rm p} = 1617$ pointed PSPC fields to radii of
$50^{\prime}$, a sky coverage of $\approx 9$\%. The mean and median
exposure times of the fields searched are $8373$~s and $5995$~s,
respectively. These should be compared with a typical RASS exposure
time of $\approx 500$~s.

We have compared observations taken during the six-month RASS (in
1990--1991) with both the small number of pointed PSPC observations
taken during the months preceding the RASS and the large number of
pointed PSPC observations taken during the $\approx 4$~yr following
the RASS, thus probing timescales ranging from months to
years. Assuming that large-amplitude X-ray outbursts decline over
periods of months to years, nearly all bright outbursts that occurred
from the $\approx 6$~months preceding the RASS to the end of the
$\approx 6$~month period of the RASS should be identifiable as
outbursts if pointed data at their positions are available.
 
\subsection{Methods and Analysis}

\subsubsection{X-ray Cross-Correlations}

To detect large-amplitude X-ray outbursts, we have cross-correlated
the {\it ROSAT All-Sky Survey Bright Source Catalogue\/} (RASSBSC;
Voges et al. 1999) with both {\it The 2nd ROSAT Source Catalog of
Pointed Observations with the PSPC\/} (ROSPSPCCAT)\footnote{See
http://wave.xray.mpe.mpg.de/rosat/rra/rospspc} and {\it The WGA
Catalog of \/\rosat\ \it Point Sources\/} (WGACAT; e.g., Angelini et
al. 2000).

We began by searching for objects that were detected in both the RASS
and pointed observations.  For each RASS source, we first determined
whether the source was located within the inner $50 ^{\prime}$ of a
pointed PSPC observation. If a pointed observation at the source
location was available, the source's offset from the center of the
field was measured, and an appropriate search radius was
determined. The positional error of \rosat\ PSPC sources increases as
the off-axis angle increases. We used a search radius which varied
linearly from $40^{\prime\prime}$ on-axis to $80^{\prime\prime}$ at an
off-axis angle of $50^{\prime}$ (Micela et al. 1996; Boese 2000;
F. G. Boese, in preparation; M. Corcoran \& S. Snowden 2002, private
communication).\footnote{Also see Figures~1a and 1b of the unpublished
report by Haberl et al. at
ftp://ftp.xray.mpe.mpg.de/rosat/catalogues/1rxp/wga\_rosatsrc.html}
For each RASS source, we then searched both the ROSPSPCCAT and WGACAT
catalogs for sources within the appropriate search radius of the RASS
position whose count rates differed from the RASS count rate by a
minimum factor of $10$. Although our survey itself is constrained to
sources with factors of variability of $20$ or greater, we manually
examined all sources with factors of variability above $10$, so as not
to miss sources variable by a factor $\geq 20$ whose catalog count
rates were slightly incorrect. Both the RASS and pointed
vignetting-corrected count rates were calculated directly from the
data using the {\sc Asterix} package (Allan \& Vallance 1995) to
ensure that all measurements were uniform.  The individual analysis of
all candidates was necessary to avoid problems such as the shadowing
of sources by the entrance window supports, catalog errors, and source
confusion.

We also searched for objects that were detected in the RASS and
undetected in pointed observations.  Specifically, we looked for RASS
sources for which a pointed PSPC observation at the source position
was available but no counterpart was found within the appropriate
error circle of the source position in either the ROSPSPCCAT or WGACAT
catalogs.  Due to sources missed by the pointed catalogs, positioning
errors, and shadowing of sources by the entrance window supports, only
a small percentage of the objects returned by this search were
actually highly variable.  In order to determine which objects were
valid candidates, pointed PSPC images of all of the sources returned
by our search were examined.  If a bright PSPC source from a pointed
observation was clearly present at the RASS source position, we
removed the source from the list of candidates.  If a dim source was
present, the count rates were calculated to search for variability by
a factor of $\geq 20$.  The distinction between undetected sources and
sources with low count rates was made using the source searching
program, {\sc PSS} (Allan 1995), in the {\sc Asterix} package.  To
determine the upper limits on undetected sources, the $95$\%
encircled-energy radius of the point spread function was calculated.
A $3\sigma$ upper limit was set by dividing the square root of the
number of counts within this radius by $0.95$ and then multiplying
this number by three. This method, which is used throughout the paper,
gives upper limits which are generally consistent with those
calculated using PSS in upper limit mode.


\section{Results}

\subsection{Sources Found in the Survey}

\subsubsection{Outbursts Detected}

Our survey has recovered all of the previously known large-amplitude
X-ray outbursts that were in a high state during the RASS and has
detected one interesting X-ray variable galaxy, SBS~$1620+545$. The
properties of the large-amplitude X-ray outbursts detected in our
survey can be found in Table~1. In addition to large-amplitude X-ray
outbursts from galaxies and AGN, four previously known cataclysmic
variables were detected in our survey (1E~$1339.8+2837$,
RE~J$2316-05$, UW Pic, and EV UMa).  We used SIMBAD to identify these
variable sources. We have therefore identified all X-ray sources with
variability amplitudes $\geq 20$ in our survey sample.

The observed variability amplitudes of the extragalactic outbursts
range from $\approx 21$ (RX~J$1420.4+5334$) to $\approx 392$
(WPVS~007).  The observed variability amplitudes are lower limits to
the true variability amplitudes, as it is possible that the outbursts
were not caught at maximum in the RASS or were not at their lowest
flux in earlier or later pointed observations.

The peak observed 0.2--2.4~keV luminosities of the detected outbursts
range from $\approx~5\times10^{41}$~erg~s$^{-1}$ to $\approx
6\times10^{43}$~erg~s$^{-1}$. These values have significant
uncertainties due to the steepness of the outburst spectra, all of
which are moderately or extremely soft (see \S 3.2 for further
discussion). The spectra of WPVS~007 and NGC~5905 are especially soft,
with hardness ratios (HR1; see Table~1) of $-0.92$ and $-0.87$,
respectively, placing them among the softest galaxies detected in the
RASS. Because the spectra of most of the outbursts are very soft,
their detection is greatly limited by high Galactic column densities.
In our survey, the maximum Galactic column density through which a
large-amplitude X-ray outburst was detected is
$3.8\times10^{20}$~cm$^{-2}$. The median Galactic column density
toward these sources is $1.4 \times10^{20}$~cm$^{-2}$.

Several of the sources detected show variability {\it during\/} the
RASS.  The count rate of NGC~5905 increases by a factor of $\approx 3$
in four days (Bade et al. 1996). Likewise, WPVS~007 shows an increase
in count rate by a factor of $\approx 2$ in two days (Grupe et
al. 1995b).  The RASS count rate of RX~J$1624.9+7554$ is observed to
vary about the mean by a factor of $\approx 2$ on the timescale of a
day (Grupe et al. 1999).

\subsubsection{SBS~$1620+545$}
Our survey has identified X-ray variability by a factor of $\geq 20$
from SBS~$1620+545$; to our knowledge, this variability has not been
reported previously.  This object has been classified as a Seyfert~2
galaxy ($z=0.0516$) by Carrasco et al. (1998) and as a Seyfert~1.9
galaxy by Veron-Cetty \& Veron (1998), but it has not been studied
further until this survey. The properties of SBS~$1620+545$ can be
found in Table~2. Figure~1 shows the RASS positional error circle
overlaid on a digitized Palomar Observatory Sky Survey (POSS) image of
the galaxy. It is clear that the variable RASS source is located in
the galaxy SBS~$1620+545$. The spectral shape of the observed X-ray
emission, however, differs from those of the large-amplitude X-ray
outbursts detected thus far. In contrast to the low hardness ratios of
the outbursts detected in our survey, the hardness ratio (HR1) of
SBS~$1620+545$ during the RASS was~$0.69$. The hard spectrum of the
observed emission, in combination with the optical classification and
relatively low factor of variability, lead us to suspect that the
variability of this object may not have been due to the same mechanism
at work in the large-amplitude X-ray outbursts that are the focus of
this study (see \S 2.1 for further discussion).  Instead, it is likely
that the variability is due to changing absorption along the line of
sight.  We note that, because the X-ray spectrum of SBS~$1620+545$ is
softer when the count rate is lower, complex absorption changes (due
to, perhaps, a partially covering or ionized absorber) would be needed
to explain the observations. It is also possible that SBS~$1620+545$
underwent an intrinsic spectral change. We consequently exclude
SBS~$1620+545$ from our $\log N$--$\log S$ analyses below, although it
is an interesting object needing further study.

\subsection{Number Counts} 

We have constructed $\log N$--$\log S$ plots of the large-amplitude
X-ray outbursts detected in our survey. We consider the cases in which
all outbursts from galaxies and AGN are included, in which only
inactive galaxies are included, and in which only active galaxies are
included (see \S 3.2.4). These plots can be found in Figure~2. Due to
the steepness of the outburst spectra, the flux measurements in these
plots are for the 0.2--2.4~keV range. If the 0.1--2.4~keV range had
been used, a large fraction of the absorption-corrected flux would be
that from 0.1--0.2~keV.  Because absorption prevents us from detecting
many photons from 0.1--0.2~keV (i.e., the spectrum from 0.1--0.2~keV
has not been well measured), it can be dangerous to extrapolate a
steep power law down to $0.1$~keV (e.g., see \S 2.3 of Brandt et
al. 1999).  Significant errors can also arise due to the poor
constraints on the true spectral shapes of these outbursting objects.
By removing the 0.1--0.2~keV range, we help to reduce the dependence
of a source's absorption-corrected flux upon its assumed low-energy
spectral shape.

The number counts in Figure~2 have been corrected for the coverage of
our survey as discussed below in \S 3.2.1 -- \S 3.2.3; they
consequently represent the number of outbursts expected to have
occurred throughout the entire sky during the time period probed by
our survey, assuming that intrinsic absorption does not prevent the
detection of such outbursts. If intrinsic absorption is significant,
our estimate of the true number of outbursts that occur in the
Universe will increase (see \S 4.1 for further discussion).

\subsubsection{Selection Effects}

It was necessary to correct for two count-rate dependent selection
effects: (1) the existence of pointed PSPC data at the position of an
outburst with a given count rate, and (2) the probability that the
exposure times of these observations are of sufficient length for the
quiescent galaxy's count rate or upper limit to be measured to the
accuracy needed to verify variability by a factor $\geq 20$. Our goal
in correcting for these factors was to determine, as a function of
absorption-corrected flux, the fraction of outbursts that were
detected by our survey, thus allowing the observed number of outbursts
at a given flux to be converted to the true number of outbursts at
that flux. This correction was applied to the differential number
counts.

Pointed PSPC data at a given outburst's position can exist for one of
two reasons; either (1) the outburst was followed-up intentionally
with the PSPC, or (2) it was observed serendipitously in a pointed
PSPC field aimed at a different target.  As we show below, X-ray
bright galaxies are more likely to have been followed-up intentionally
with the PSPC than X-ray faint galaxies. In addition, the probability
that the pointed count rate or upper limit of a galaxy caught
serendipitously in a PSPC field can be measured to the accuracy needed
to verify variability by a factor of $\geq 20$ is higher for bright
galaxies than for faint galaxies. Consequently, outbursts with high
fluxes are more likely to have been discovered than outbursts with low
fluxes. The fraction of the sky, $F$, in which outbursts with a given
flux would have been discovered was determined based on these
selection effects. Here and hereafter, we consistently neglect the
region of the sky from Galactic latitudes of $-30^{\circ}$ to
$+30^{\circ}$ in our calculations (see \S 2.1). To carry out the
calculation of $F$, we assume that the outbursting galaxies are
distributed isotropically, a valid assumption for extragalactic
sources.

The fraction $F$ is the sum of two components.  The first, $F_{\rm
f}$, is the fraction of the sky in which a RASS source with a given
flux would have been likely to be followed-up intentionally with the
PSPC, thus allowing the outburst to be discovered.  The second,
$F_{\rm s}$, is the fraction of the sky in which a RASS source would
not have been intentionally followed-up, but the outburst would have
been located serendipitously in a PSPC field aimed at a different
target. We note that these corrections are intended to determine the
flux-dependent fraction of the sky in which {\it any\/}
large-amplitude X-ray outburst from a galaxy or AGN, not only the five
that were found in our survey, would have been detected.  Because the
corrections are not being made for the specific outbursts detected,
but rather to account for those outbursts from galaxies and AGN that
were {\it not\/} detected, the median photon index ($\Gamma = 4.0$) of
the large-amplitude X-ray outbursts in our survey is used below in the
calculation of the correction factor.

\subsubsection{Probability of Intentional Follow-up}

In order to estimate $F_{\rm f}$, we first quantified the effect that
an extragalactic source's RASS flux had on its probability of being
followed-up with the PSPC.  To do so, we used Version 2.0 of {\it The
Hamburg/RASS Catalog of Optical Identifications\/} (Bade et al. 1998).
This catalog contains optical identifications for 4665 RASSBSC sources
in the extragalactic northern sky, and it was published after the PSPC
ceased operation. Objects in this catalog were classified using the
objective prism and direct plates taken by the Hamburg Quasar Survey
(HQS).  Although $\approx 20$\% of the RASSBSC sources in this catalog
are listed as either ``unidentified'' or ``empty'', the majority of
such sources are believed to be extragalactic; we therefore consider
all ``unidentified'' and ``empty'' sources to be extragalactic.
Consequently, the Hamburg/RASS catalog allowed us to identify all
extragalactic RASSBSC sources in the region of sky covered by the HQS.
By cross-correlating the positions of the extragalactic Hamburg/RASS
sources with the central positions of the pointed PSPC observations
used in our survey, we were able to determine the percentage of
extragalactic RASSBSC sources with a given count rate that were
intentionally followed-up with pointed PSPC observations.  This
function, $P(R_{\rm o})$, where $R_{\rm o}$ is the observed RASS count
rate, can be found in Figure~3.  Because large-amplitude X-ray
outbursts tend to be soft, we also measured $P(R_{\rm o})$ for only
the extragalactic sources in the Hamburg/RASS catalog with negative
hardness ratios (HR1).  This function does not differ significantly
from that for all extragalactic Hamburg/RASS sources.

To determine $F_{\rm f}$, we considered $N_{\rm s} = 92$ approximately
equally spaced points on the high Galactic latitude survey sky.  At
each point, the Portable Interactive Multi-Mission Simulator (PIMMS;
Mukai 2000) was used to calculate the expected RASS count rate,
$R_{\rm n}$, of an outburst with a given flux, based on the Galactic
column density toward that position. The probability that an outburst
with a given flux would have been followed up at one of these points
is given by the value of the function $P(R_{\rm n})$ at this point,
and it is taken to be equivalent to the probability that it would have
been followed-up if it were located in the area of the sky, $A =
A_{\rm survey}/92$ nearest this point, where $A_{\rm survey} =
20627$~deg$^{2}$ is the area of the sky outside of the Galactic plane
as defined in \S 2.1.  The fraction of the sky sampled by our survey
in which the outburst would have been followed-up and subsequently
discovered is then given by:

\begin{equation}
F_{\rm f} = \sum^{N_{\rm s}}_{n=1} {{P(R_{\rm n})}\over N_{\rm s}}
\label{eqno1}
\end{equation}

\subsubsection{Probability of Serendipitous Detection}

To calculate $F_{\rm s}$ at each of the fluxes at which an outbursting
galaxy was detected, we first found $A(R_{\rm \ell})$, the area
of a given PSPC field in which a limiting on-axis count rate, $R_{\rm
\ell}$, would have been detected. To do so, a vignetting-corrected
limiting count rate was calculated at fifteen off-axis angle intervals
of $3^{\prime}\!\!.33$. At each off-axis angle interval, we measured
the number of background counts within the $95$\% encircled-energy
radius of the appropriate point spread function at that off-axis
angle. The $3\sigma$ limiting count rate was then calculated as
described in \S 2.2.1. To normalize this function, we then divided the
minimum detectable count rate, $R_{\rm
\ell}$, by the central minimum detectable count rate, $R_{\rm C}$,
converting $A(R_{\rm \ell})$ to $A({R_{\rm \ell}/R_{\rm C}})$ (The
calculation of $R_{\rm C}$ is discussed below).  We found this
relationship for several PSPC observations covering a range of \rosat\
exposure times.  Although this function is slightly different for each
PSPC observation, its overall shape and limits are largely consistent
between the observations we have used here, once the background level
is properly accounted for (recall also that PSPC fields with unusually high
backgrounds were removed from the survey in \S 2.1.)  Because of this
consistency, we were able to average the results of several PSPC
observations to determine an average \abar(${R_{\rm
\ell}/ R_{\rm C}})$ function. To create an observation-specific
function
\abar$(R_{\rm \ell})$ for each PSPC field searched, we scaled the
function \abar$({R_{\rm \ell}/R_{\rm C}})$ by the central limiting
count rate, $R_{\rm C}$, of each observation.  The equation for the
central limiting count rate as a function of exposure time for
\rosat\ PSPC observations was taken from \S 10.1 of the \rosat\ Mission
Description (Appendix F). We made the assumption that a minimum of
$10$ counts is required to constitute a detection at the center of a
PSPC field. The background count rate for each field was taken from
either the ROSPSPCAT or WGACAT catalogs; the appropriate corrections
were made to convert the background count rates to the 0.2--2.4~keV
energy band. The calculation of the central limiting count rate was
performed for a signal-to-noise of 3 and an enclosed source count
fraction of $0.95$.

To calculate the area of a given PSPC field in which the quiescent
count rate or upper limit of an outbursting galaxy would have been
measurable to the accuracy needed to demonstrate variability by a
factor $\geq 20$, we measured the value of \abar$(R_{\rm \ell})$ at
$R_{\rm \ell} = R_{\rm n}/20$. Recall from \S 3.2.2 that $R_{\rm n}$
is the expected RASS count rate at a given position for an outburst
with a given flux, and therefore $R_{\rm n}/20$ is the limiting count
rate at that position above which large-amplitude variability cannot
be demonstrated. The area of a given pointed PSPC field in which an
outburst would have been discovered if and only if it were detected
serendipitously is given by the area of that field in which the count
rate $R_{\rm n}/20$ would have been measurable, ${\bar{A}{(R_{\rm
n}/20)}}$, times the probability that the outburst would not have been
intentionally followed-up.  The value, $F_{\rm s}$, is then given by
scaling this area to the area of the sky probed by the survey, $A_{\rm
survey}$, and summing over all $N_{\rm p} = 1617$ PSPC fields used in
our survey:

\begin{equation}
F_{\rm s} = \sum^{N_{\rm p}}_{n=1} {{\bar{A}(R_{\rm
n}/20)}\over{A_{\rm survey}}}\times[1 - P(R_{\rm n})]
\label{equao2}
\end{equation}

The total fraction of the survey sky in which a given large-amplitude
X-ray outburst would have been detected, $F$, is thus given by:

\begin{equation}
F = F_{\rm f} + F_{\rm s} = \sum^{N_{\rm s}}_{n=1} {{P(R_{\rm
n})}\over N_{\rm s}} + \sum^{N_{\rm p}}_{n=1} {{\bar{A}(R_{\rm
n}/20)}\over{A_{\rm survey}}}\times[1 - P(R_{\rm n})]
\label{equao3}
\end{equation}

Because the region of the sky from Galactic latitudes of $-30^{\circ}$
to $+30^{\circ}$ constitutes half of the area of the sky, the final
fraction, $F$, was divided by 2 to scale the results to the entire
sky. The value of $F$ used to correct the differential number count of
each outburst detected in our survey is given in Table~1. While we
recognize that correction for these complicated systematic effects is
challenging, we believe that the method described above provides the
best practical correction.

\subsubsection{$\log N$--$\log S$ Plots}

To construct the corrected $\log N$--$\log S$ plots, we first divided
each outburst's differential number count, 1, by the fraction of the
survey area in which an outburst with that flux would have been
discovered, $F$.  We then converted from differential number counts,
$N(S)$, to cumulative number counts, $N(>S)$.  The error bars on the
corrected number counts have been calculated using small number
statistics as outlined in Gehrels (1986), i.e. it was assigned to be
the small-number statistical error on 1, divided by $F$.  When
converting from differential to cumulative number counts, these errors
were added in quadrature.  We note that the points in Figure~2 lie
roughly on straight lines. The reason for this behavior is that the
the correction function, $F$, as a function of flux, is nearly linear
in $\log$--$\log$ space. The corrected $\log N$ value for each point
is determined entirely by this correction function.  As such, the
error bars on the number counts should not be interpreted as standard
statistical errors, but are instead representative of the uncertainty
introduced by the correction itself (i.e., as the correction to the
differential number counts increases, the error bars on the corrected
number counts increase).  We believe that this method provides the
most reasonable estimation possible of the complicated errors
associated with this work.

In Figure~2a, we plot number counts derived using all of the
large-amplitude X-ray outbursts that occurred during the RASS. In
Figure~2b, we plot only those outbursts that originated from inactive
galaxies, for which the stellar tidal disruption outburst scenario is
most likely to be applicable. In Figure~2c, we plot those outbursts
from active galaxies, for which accretion-disk instabilities may be
responsible. We have determined for each of the above cases both the
best-fit power laws [$N(>S) \propto S^{\rm \alpha}$] to the data
points as well as the best-fit power laws for fixed power-law indices
of $\alpha = -3/2$; a power law index of $-3/2$ is expected for a
sample of uniformly distributed sources in Euclidean space. The fit
parameters can be found in Table~3.

The best-fit $\alpha$ values for the cases in which all outbursts are
considered and in which only inactive galaxies are considered are
approximately $-1$.  The flatness of this parameter may be due to
small-number statistical fluctuations or to the superposition of more
than one population. Because of the small number of detected sources
and the likely heterogeneity of the sample, it is difficult to rule
out any of the number count fits; we consequently consider all cases
when placing constraints on the frequency of large-amplitude X-ray
outbursts in the local Universe.

\subsection{Survey Volume}

To determine the characteristic outburst flux limit and volume to
which our survey is complete, we converted the function $A(R_{\rm
\ell}/R_{\rm C})$, the area of a given PSPC field in which a limiting
on-axis count rate would have been detected, to its differential form.
The resultant function, $A_{\rm diff}(R_{\rm \ell}/R_{\rm C})$, gives
the area of a PSPC field with a given limiting on-axis count
rate. After scaling this function to each PSPC observation, we summed
it over all PSPC fields used in this survey.  This summed function
gives the area of our survey with a given limiting
absorption-corrected count rate and has a peak at 0.031~cts~s$^{-1}$.
We adopt this peak value as the characteristic 0.2--2.4~keV
absorption-corrected PSPC count rate to which our survey is complete.
Using PIMMS, we have converted this count rate to a 0.2--2.4~keV
unabsorbed PSPC flux, $1 \times 10^{-13}$~erg~cm$^{-2}$~s$^{-1}$. The
above conversion was performed using the median spectral index
($\Gamma = 4.0$) of the large-amplitude X-ray outbursts in our
survey. Because we have defined large-amplitude X-ray outbursts to be
sources whose luminosity drops by a minimum factor of 20 between the
RASS and pointed PSPC observations, the characteristic
absorption-corrected RASS flux to which we are complete is equal to 20
times the characteristic absorption-corrected PSPC flux, or $2 \times
10^{-12}$~erg~cm$^{-2}$~s$^{-1}$.

The median 0.2--2.4~keV outburst (RASS) luminosity of the outbursts
detected in our survey is $L_{\rm outburst} = 2.8 \times
10^{43}$~erg~s$^{-1}$.  The characteristic distance to which we are
complete is therefore $\approx 342$~Mpc, corresponding to $z \approx
0.091$. Our survey has consequently covered a volume of $V_{\rm
survey} \approx 1.68 \times 10^{8}$~Mpc$^{3}$. Because large-amplitude
X-ray outbursts decay over a period of months to years, the majority
of outbursts that flared from the six months preceding the RASS to the
end of the six-month RASS, and for which pointed PSPC observations
exist, should have been found by our survey.  We therefore consider
our observed rates to be those for a period of approximately one year.
To determine the effect of the X-ray ``$K$-correction'' on our
completeness, we simulated redshifted spectra with the median
properties listed above using XSPEC (Arnaud 1996).  The $K$-correction
causes the observed count rate to drop by $\approx 27\%$ at our
completeness limit of $z=0.091$; we consequently neglect this effect
because it is significantly smaller than the geometric dilution
factor.

 
\section{Discussion and Conclusions} 

\subsection{Constraints}

To place constraints on the rate of large-amplitude X-ray outbursts in
the local Universe, we consider first the case in which all outbursts
from galaxies and AGN have been included in the $\log N$--$\log S$ fit
and for which the power-law index is fixed at $\alpha = -3/2$.
Although it is possible that the outbursts from active and inactive
galaxies arise from different mechanisms, these events are still
poorly understood, making such a calculation worthwhile.  The outburst
rates derived from all fits can be found in Table~3.  For the fit
described above, $\approx 42$ outbursts should have occurred
throughout the entire sky down to our characteristic completeness flux
of $2 \times 10^{-12}$~erg~cm$^{-2}$~s$^{-1}$. We assume a space
density, $1.4 \times 10^{-2}$~Mpc$^{-3}$, that is the sum of the
inactive galaxy space density and the active galaxy space density. The
inactive galaxy space density of $1.35 \times 10^{-2}$~Mpc$^{-3}$ is
the sum of a spiral galaxy space density of $1 \times
10^{-2}$~Mpc$^{-3}$ (e.g., de Jong 1996) and an E+S0 space density of
$3.5\times 10^{-3}$~Mpc$^{-3}$ (e.g., Magorrian \& Tremaine 1999). We
assume an active galaxy space density of $5 \times 10^{-4}$~Mpc$^{-3}$
(e.g., Peterson 1997).  The outburst rate for all galaxies and AGN is
then $1.8 \times 10^{-5}$~galaxy$^{-1}$~yr$^{-1}$. If intrinsic
absorption is important, this rate will be higher; this effect is
discussed in detail below for the case in which only outbursts from
inactive galaxies are included in the $\log N$--$\log S$ fit.

When only the outbursts from inactive galaxies are considered, we
calculate an outburst rate of $9.1 \times
10^{-6}$~galaxy$^{-1}$~yr$^{-1}$. This rate corresponds to a timescale
of $1.1\times 10^{5}$~yrs between outbursts for a given inactive
galaxy.  Because this rate lies just below the predicted maximum rates
for tidal disruption events in inactive galaxies (see \S 1), this
result provides additional support for the hypothesis that
large-amplitude X-ray outbursts from inactive galaxies are the result
of stellar tidal disruptions by supermassive black holes.  In
addition, this rate implies that down to 0.02~cts~s$^{-1}$, the
typical count rate of a source is the \rosat\ All-Sky Survey Faint
Source Catalog (RASSFSC)\footnote{See
http://www.xray.mpe.mpg.de/rosat/survey/rass-fsc/}, $\approx 2000$
outbursts from inactive galaxies should have occurred during the RASS.
This result is in agreement with the prediction of Sembay \& West
(1993; see \S 1). We note that if intrinsic absorption prevents
large-amplitude X-ray outbursts from being discovered, this rate will
increase.  This effect was investigated in \S 5 of Sembay \& West
(1993); they estimate that X-ray outbursts could be detected from only
$\approx 1/2$ of all spiral galaxies.  If we assume, from the
respective number densities of spiral and E+S0 galaxies, that $\approx
40$\% of the inactive galaxies sampled by our survey are such roughly
edge-on spirals, an outburst would be detectable from only $f \approx
60$\% of the galaxies in our sample. Consequently, the outburst rate
would rise to $1.5 \times 10^{-5}$~galaxy$^{-1}$~yr$^{-1}$.  Given the
uncertainty on $f$, however, we perform all calculations under the
assumption that $f$ is equal to unity.

If we consider only the two active galaxies in our sample for the case
in which $\alpha = -3/2$, we obtain an active galaxy outburst rate of
$8.5\times10^{-4}$~galaxy$^{-1}$~yr$^{-1}$. This rate is substantially
higher than that for inactive galaxies. Because of the theoretical
uncertainties associated with the tidal disruption of a star in an
active galaxy, predictions of the rate of such events have not yet
been made.  Consequently, it is difficult to determine whether the
derived outburst rate for active galaxies is consistent with the
stellar tidal disruption scenario.  The dynamics of a stellar tidal
disruption event are expected to differ for a galaxy in which an
accretion disk is present. It is thought that interactions between a
star and the accretion disk of an active galaxy may allow the star to
lose momentum and energy and reach a radius at which tidal disruption
could occur (e.g., Syer, Clarke, \& Rees 1991; Armitage, Zurek, \&
Davies 1996); the effect that this interaction would have on the rate
of disruption is not yet clear.  The high apparent rate of outbursts
from AGN, however, suggests that at least some of these outbursts may
be due to another mechanism, such as accretion-disk instabilities
(e.g., Siemiginowska et al. 1996; Burderi et al. 1998), which could
lead to variations in the soft excess.

\subsection{Predictions for Future Work}

In order to place tighter constraints on the outburst rates of both
active and inactive galaxies, additional outbursts need to be
detected.  It would also be useful if the outbursts could be
followed-up in order to (1) determine their decay rates, (2) measure
their X-ray spectra, and (3) look for associated spectroscopic
signatures in other bands.  In the future, cross-correlations of
\rosat, \xmm, and \chandra\/ data may provide a means by which to
identify further outbursts.  Assuming that \xmm\/ operates for
$\approx 10$~yrs at 70\% efficiency, taking observations with a mean
length of 40~ks, it will produce $\approx 5500$ observations.  If
$\approx 1000$ of these observations overlap and $\approx 30$\% are in
the Galactic plane, \xmm\/ should produce $\approx 3200$ distinct
extragalactic observations.  Of these observations, $\approx 35$\%
should have column densities $\geq 3.8 \times 10^{20}$~cm$^{-2}$, the
highest column density through which an outburst in our survey was
detected.  Consequently, $\approx 2100$ observations will be useful
for identifying large-amplitude X-ray outbursts.  The EPIC PN camera,
which is the \xmm\/ instrument most sensitive to soft X-rays, has a
field of view of $\approx 718$~arcmin$^{2}$.  Approximately 1/2 of
this area, $\approx 359$~arcmin$^{2}$, is of sufficient sensitivity to
detect these outbursting objects in quiescence for a 40~ks exposure.
Consequently, \xmm\/ will cover $\approx 0.5$ \% of the extragalactic,
low Galactic column density sky.  Cross correlation of these data with
the RASS catalogs should allow the identification of all RASS
outbursts with outburst count rates $\geq 0.02$~cts~s$^{-1}$, the
typical count rate of a source in the \rosat\/ All-Sky Survey Faint
Source Catalogue (RASSFSC).\footnote{See
http://www.xray.mpe.mpg.de/rosat/survey/rass-fsc/} A count rate of
0.02~cts~s$^{-1}$ corresponds to an outburst flux of $\approx 9\times
10^{-14}$~erg~cm$^{-2}$~s$^{-1}$.  Consequently, cross-correlation
between RASS and \xmm\/ observations should result in the detection of
$\approx 22$ RASS outbursts. We have carried out a similar calculation
for \chandra\/ and find that, if \chandra\/ were highly sensitive in
the 0.2--0.3~keV band, it could identify $\approx 13$ RASS outbursts.
The back-illuminated and front-illuminated ACIS CCD chips, however,
are sensitive down to $\approx 0.3$~keV and $\approx 0.5$~keV,
respectively.  Because the flux of a typical large-amplitude X-ray
outburst drops by a factor of $\approx 10$ when the flux from
0.2--0.3~keV is excluded, it will be significantly more difficult to
prove variability by a factor of 20 without sensitivity in this energy
range.  Consequently, only a handful of RASS outbursts could be
expected to be detected through cross-correlations of \rosat\/ and
\chandra\/ data.

Any correlation between all-sky survey and pointed observations, or
pointed observations and pointed observations, will be limited by the
complicated selection effects discussed in this paper, and perhaps
others as well. As such, all-sky monitors, or a new sensitive all-sky
survey, would provide the best means by which to detect and study
additional outbursts in a uniform manner.  Missions such as \lobster\
(e.g., Parmar 2001), \maxi\ (e.g., Mihara et al. 2000),
\rosita\footnote{See http://wave.xray.mpe.mpg.de/rosita}, and \swift\
(e.g., Gehrels 2000), however, probably do not have the combined
spatial resolution and sensitivity to soft X-rays needed to identify
additional outbursts of this type (see Grupe 2002). As such, a new
sensitive soft X-ray all-sky survey or all-sky monitor is needed.
Based on our fixed-$\alpha$ $\log N$--$\log S$ fit for all outbursts
from galaxies and AGN, we find that down to a count rate of
$0.02$~cts~s$^{-1}$, the typical count rate of a source in the
RASSFSC, $\approx 4400$ large-amplitude X-ray outbursts should have
occurred during the RASS.  Although the Galactic column density toward
$\approx 65$\% of these sources is higher than the maximum Galactic
column through which one of the outbursts in our sample was detected,
$\approx 1540$ outbursts should have occurred in regions of sky where
the Galactic column density is sufficiently small. In order to detect
and study these events, a second all-sky survey sensitive in the
0.1--2.4~keV range is needed.  Such a survey should be deep enough to
detect sources with fluxes of $\approx 5 \times
10^{-15}$~erg~cm$^{-2}$~s$^{-1}$, a factor of 20 below the flux of an
outburst with a 0.1--2.4 keV RASS count rate of
$0.02$~cts~s$^{-1}$. Cross-correlation of the RASS with a new all-sky
survey would remove many of the complicated selection effects
considered above and would provide an excellent means to increase
greatly the number of known outbursts of this type, allowing tighter
constraints to be placed on the rate of these outbursts in the
Universe. A sensitive soft X-ray all-sky monitor would provide an
excellent means to detect newly outbursting objects and study them as
their fluxes decline. Observations of possible tidal disruption events
and accretion-disk instabilities in other wavelength bands will also
be of use in understanding better these events.


\acknowledgments
We thank L. Angelini for providing us with the list of PSPC
observations used to construct the WGACAT catalog.  We also thank
R. Ciardullo, M. Corcoran, S. Sigurdsson, and S. Snowden for helpful
discussions. We thank an anonymous referee for comments. We gratefully
acknowledge financial support from NSF CAREER award AST-9983783 (JLD,
WNB) and the Penn State President's Fund for Undergraduate Research
(JLD). Some of the spectra presented here were obtained with the
Hobby-Eberly Telescope, which is a joint project of the University of
Texas at Austin, the Pennsylvania State University, Stanford
University, Ludwig-Maximillians-Universit\"at M\"unchen, and
Georg-August-Universit\"at G\"ottingen. The Marcario Low-Resolution
Spectrograph (LRS) is a joint project of the Hobby-Eberly Telescope
partnership and the Instituto de Astronom\'ia de la Universidad
Nacional Aut\'onoma de M\'exico.


\newpage

\appendix
{
\centerline{\bf APPENDIX}
\section{Flaring X-ray Sources Near the Galactic Plane}

Although sources in the Galactic plane and sources that varied between
pointed observations are not part of our survey (see \S 2.1), we
extended our search for interesting variable sources to these samples.
We obtained optical spectra for several variable objects, all of which
are stars.  Optical spectroscopy of candidate flaring X-ray sources
near the Galactic plane was carried out with three different
telescopes on separate occasions, as described in
Table~A-1. Figure~A-1 shows images of the four fields observed. In
each image we mark the PSPC error circle, and we identify the objects
that we observed spectroscopically. In Figure~A-2 we show a montage of
the spectra of these objects, all of which turned out to be stars.  It
is noteworthy that in two of the four fields we found late M stars
(one a dMe star) within or very near the PSPC error circle, which
suggests that these are the flaring objects. In the other two fields
the optical objects within or just around the PSPC error circle turned
out to be F or G stars, which we regard as unlikely counterparts of
the flaring X-ray sources.  }



\newpage

\begin{figure}
\epsscale{0.5}
\plotone{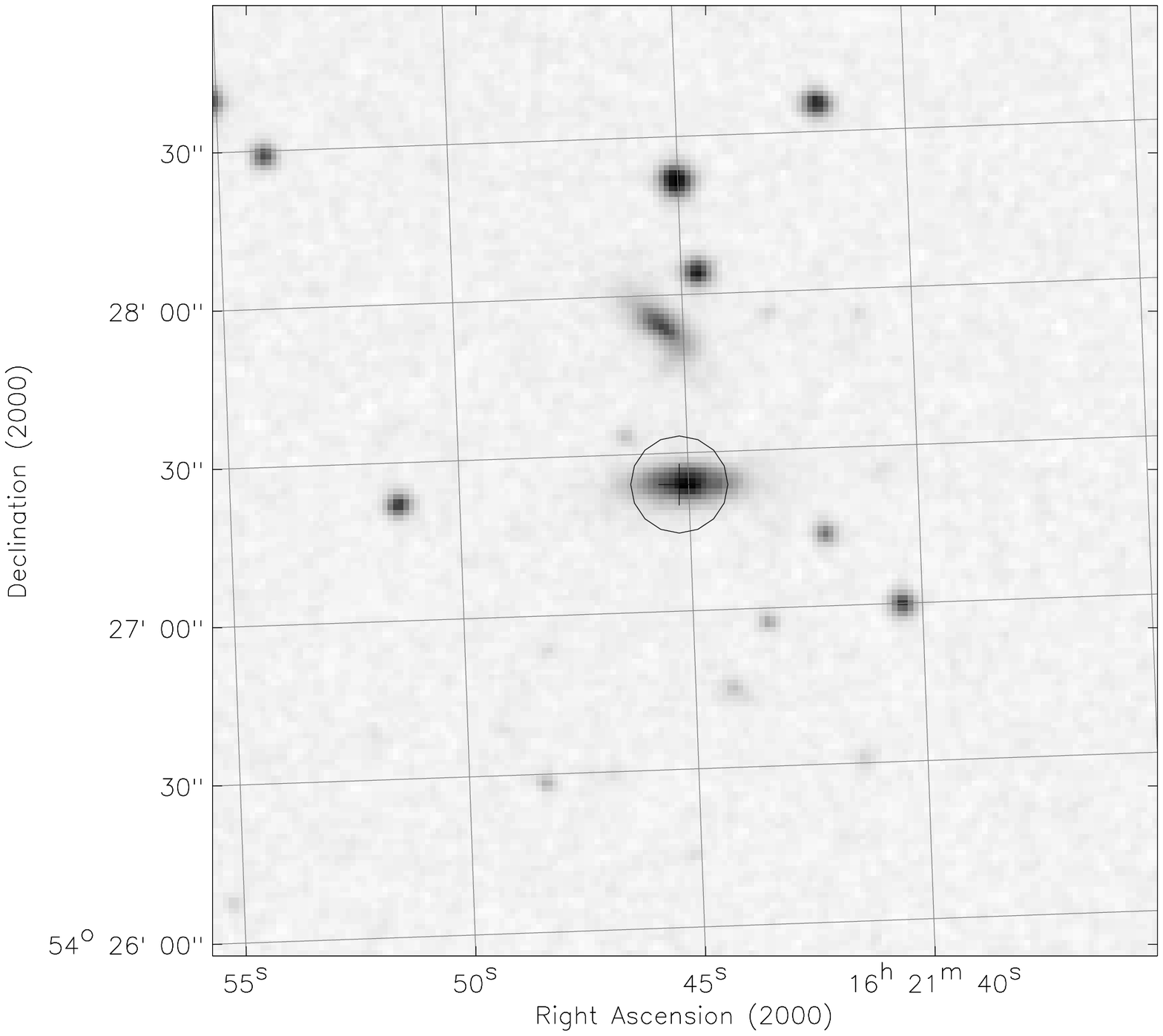}

\caption{POSS image of the galaxy SBS $1620+545$ with the 9$^{\prime\prime}$
radius RASS positional error circle overlaid.}

\end{figure}


\newpage

\begin{figure}
\epsscale{1.0}
\plotone{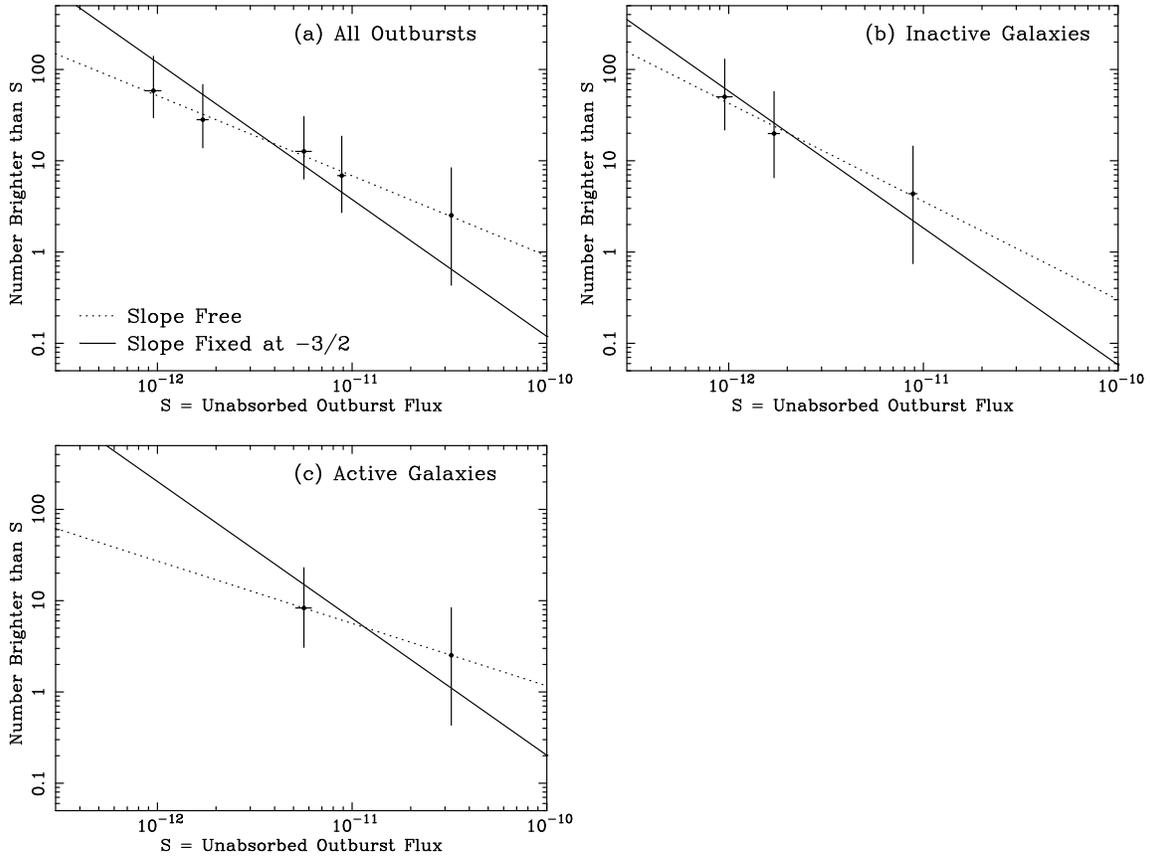}

\caption{Corrected $\log N$--$\log S$ plots for X-ray outbursts
 during the RASS. All outbursts from galaxies and AGN are plotted in
 Figure~2a.  In Figure~2b, only the inactive galaxies are plotted.
 The active galaxies are plotted in Figure~2c.  We have separated the
 active and inactive galaxies because it is possible that a different
 mechanism is responsible for the large-amplitude X-ray outbursts
 arising from these two groups of objects.}

\end{figure}


\newpage

\begin{figure}
\epsscale{0.75}
\plotone{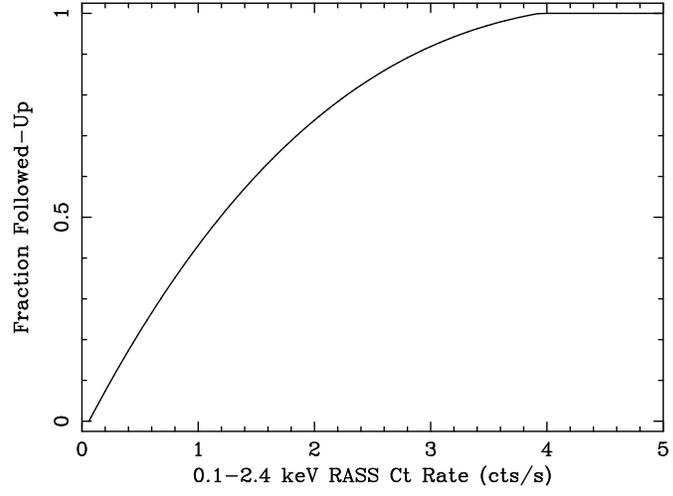}

\caption{Fit to the fraction of extragalactic Hamburg/RASS sources
that were followed-up with pointed PSPC observations as a function of
their RASS count rate.}

\end{figure}


\newpage
\thispagestyle{empty}
\begin{deluxetable}{lllllllllllllll}

\tablenum{1}
\tabletypesize{\tiny}
\rotate
\tablewidth{0pt}
\tablecaption {Large-Amplitude X-ray Outbursts in the Survey}

\tablehead{
\colhead{Name}                                &   
\colhead{$\alpha_{2000}$}                     &
\colhead{$\delta_{2000}$}                     &    
\colhead{$N_{\rm H}$\tablenotemark{a}}        &
\colhead{$z$}                                 &
\colhead{Amp$_{\rm var}$\tablenotemark{b}}    &
\colhead{Phase}                               &    
\colhead{Date}                                &                            
\colhead{CR\tablenotemark{c}}                 &
\colhead{HR1\tablenotemark{d}}                &
\colhead{$\Gamma$\tablenotemark{e}}           &
\colhead{$F_{\rm x}$\tablenotemark{f}}        &
\colhead{$L_{\rm x}$\tablenotemark{g}}        &
\colhead{$F$\tablenotemark{h}}                &
\colhead{Ref.\tablenotemark{i}}               
}

\startdata

WPVS 007          & $00\ 39\ 15.8$  & $-51\ 17\ 03$    & 2.6  & 0.0288 &392    & RASS    & 1990 Nov 10--12 & $1.02 \pm 0.07$     & $-0.92^{+0.09}_{-0.10}$    & 8.3           & $5.66  \pm 0.53$    & 0.859    &0.17     & 1   \\
                  &                 &                  &      &        &       & Pointed & 1993 Nov 11--13 & $0.0026 \pm 0.0013$ & $-0.59^{+0.27}_{-0.31}$    & 4.2           & $0.028 \pm 0.012$   & 0.0019   &         &     \\

IC 3599           & $12\ 37\ 41.2$  & $+26\ 42\ 27$    & 1.3  & 0.0215 &225    & RASS    & 1990 Dec 10--11 & $5.18 \pm 0.11$     & $-0.52^{+0.02}_{-0.02}$    & 3.1           & $32.33 \pm 0.76$   & 2.77      &0.40     & 2,3 \\
                  &                 &                  &      &        &       & Pointed & 1993 Jun 17     & $0.023 \pm 0.004$   & $-0.87^{+0.17}_{-0.19}$    & 4.1           & $0.093 \pm 0.018$  & 0.0080    &         &     \\

RX J$1420.4+5334$ & $14\ 20\ 24.4$  & $+53\ 34\ 12$    & 1.2  & 0.147  &$> 21$ & RASS    & 1990 Dec 5--8   & $0.31 \pm 0.02$     & $-0.88^{+0.09}_{-0.10}$    & $5.8 \pm 0.6$ & $0.95 \pm 0.09$    & 3.21      &0.03     & 4   \\
                  &                 &                  &      &        &       & Pointed & 1990 Jul 19--23 & $\leq 0.016$        & $\cdots$                   & $\cdots$      &                    &           &         &     \\

NGC 5905          & $15\ 15\ 23.2$  & $+55\ 31\ 05$    & 1.4  & 0.0126 &45     & RASS    & 1990 Jul 11--16 & $0.33 \pm 0.02$     & $-0.87^{+0.07}_{-0.08}$    & $4.0 \pm 0.4$ & $1.71 \pm 0.11$    & 0.0509    &0.06     & 5   \\
                  &                 &                  &      &        &       & Pointed & 1993 Jul 18     & $0.0073 \pm 0.0011$ & $-0.11^{+0.14}_{-0.14}$    & $2.4 \pm 0.7$ & $0.045 \pm 0.009$  & 0.0014    &         &     \\

RX J$1624.9+7554$ & $16\ 24\ 56.5$  & $+75\ 54\ 56$    & 3.8  & 0.0636 &$> 42$ & RASS    & 1990 Oct 7--15  & $0.54 \pm 0.02$     & $-0.08^{+0.04}_{-0.04}$    & $3.3 \pm 0.2$ & $8.85 \pm 0.43$    & 6.24      &0.23     & 6   \\
                  &                 &                  &      &        &       & Pointed & 1992 Jan 13     & $\leq 0.013$        & $\cdots$                   & $\cdots$      &                    &           &         &     \\
    
\enddata

\tablecomments {Units of right ascension are hours, minutes, and seconds, and units of declination are degrees, arcminutes, and arcseconds.} 
\tablenotetext{a} {Galactic column density in units of $10^{20}$ cm$^{-2}$.}
\tablenotetext{b} {Observed amplitude of count-rate variability in the 0.1--2.4 keV band.}
\tablenotetext{c} {Count rate in the $0.1$--$2.4$~keV energy band in cts~s$^{-1}$.}
\tablenotetext{d} {HR1 = (H$-$S)/(H$+$S) where S is the $0.1$--$0.4$ keV count rate and H is the $0.4$--$2.4$ keV count rate.} 
\tablenotetext{e} {The photon indices of the outbursting galaxies were taken from the respective papers cited and are the best
fits for power-law models with the column density fixed at the Galactic value. For WPVS~007, the pointed-phase photon index was
estimated from the hardness ratio.}
\tablenotetext{f} {Observed-frame absorption-corrected X-ray flux from $0.2$--$2.4$~keV in units of $10^{-12}~$erg~cm$^{-2}$~s$^{-1}$. See \S 3.2 for discussion.}
\tablenotetext{g} {Observed-frame absorption-corrected X-ray luminosity from $0.2$--$2.4$~keV in units of $10^{43}~$erg~s$^{-1}$. See \S 3.2 for discussion.}
\tablenotetext{h} {$F$ is the correction factor applied to each outburst's differential number count.  See \S3.2.3 for discussion.}
\tablenotetext{i} {References are (1) Grupe et al. 1995b, (2) Brandt et al. 1995, (3) Grupe et al. 1995a, (4) Greiner et al. 2000, 
                  (5) Bade et al. 1996, (6) Grupe et al. 1999.}

\end{deluxetable}

\clearpage


\newpage
\thispagestyle{empty}
\begin{deluxetable}{lllllllllllll}
\tablenum{2}
\tabletypesize{\tiny}
\rotate
\tablewidth{0pt}
\tablecaption {The X-ray Variable Galaxy SBS 1620+545}

\tablehead{
\colhead{Name}                                &   
\colhead{$\alpha_{2000}$}                     &
\colhead{$\delta_{2000}$}                     &    
\colhead{$N_{\rm H}$\tablenotemark{a}}        &
\colhead{$z$}                                 &
\colhead{Amp$_{\rm var}$\tablenotemark{b}}    &
\colhead{Phase}                               &    
\colhead{Date}                                &                            
\colhead{CR\tablenotemark{c}}                 &
\colhead{HR1\tablenotemark{d}}                &
\colhead{$\Gamma$\tablenotemark{e}}           &
\colhead{$F_{\rm x}$\tablenotemark{f}}        &
\colhead{$L_{\rm x}$\tablenotemark{g}}                
}

\startdata

SBS $1620+545$     & $16\ 21\ 45.1$   & $+54\ 27\ 24$      & 1.9 & 0.0516  &20 & RASS      & 1991 Jan 14--22  & $0.15 \pm 0.01$       & $+0.69^{+0.123}_{-0.131}$    & 0.96      &$2.185 \pm 0.280$      & 1.11    \\
                   &                  &                    &     &         &   & Pointed   & 1993 Sep 22--24  & $0.0074 \pm 0.0024$   & $+0.31^{+0.192}_{-0.199}$    & $\cdots$  &$0.085 \pm 0.084$      & 0.043   \\
      
\enddata

\tablecomments{Units of right ascension are hours, minutes, and seconds, and units of declination are degrees, arcminutes, and arcseconds.} 
\tablenotetext{a} {Galactic column density in units of $10^{20}$ cm$^{-2}$.}
\tablenotetext{b} {Observed amplitude of count-rate variability in the 0.1--2.4 keV band.}
\tablenotetext{c} {Count rate in the $0.1$--$2.4$~keV energy band in cts~s$^{-1}$.}
\tablenotetext{d} {HR1 = (H$-$S)/(H$+$S) where S is the $0.1$--$0.4$ keV count rate and H is the $0.4$--$2.4$ keV count rate.} 
\tablenotetext{e} {The photon index was estimated from the hardness ratio.}
\tablenotetext{f} {Observed-frame absorption-corrected X-ray flux from $0.2$--$2.4$~keV in units of $10^{-12}$erg~cm$^{-2}$~s$^{-1}$. See \S 3.2 for discussion.}
\tablenotetext{g} {Observed-frame absorption-corrected X-ray luminosity from $0.2$--$2.4$~keV in units of $10^{43}~$erg~s$^{-1}$. See \S 3.2 for discussion.}

\end{deluxetable}

\clearpage


\newpage
\thispagestyle{empty}
\begin{deluxetable}{lllc}
\tablenum{3}
\tabletypesize{\small}
\rotate
\tablewidth{0pt}
\tablecaption {Number Count Fits: $N(>S) = A({S\over 10^{-12}})^{\alpha}$}

\tablehead{
\colhead{}                                      &
\colhead{}                                      &
\colhead{}                                      &
\colhead{Outbursts}                             \\
\colhead{Parameters}                            &   
\colhead{$A$\tablenotemark{a}}                  &
\colhead{$\alpha$\tablenotemark{a}}             &    
\colhead{Galaxy$^{-1}$~Year$^{-1}$}                                      
                      
}

\startdata

All outbursts, $\alpha$ free            & 51.7$^{+97.9}_{-33.84}$     & $-0.88\pm 0.60$      & $1.2 \times 10^{-5}$  \\
All outbursts, $\alpha$  fixed          & 118.9$^{+119.4}_{-59.6}$    & $-1.50 $             & $1.8 \times 10^{-5}$  \\
Inactive galaxies, $\alpha$  free       & 42.7$^{+111.8.0}_{-30.9}$   & $-1.08\pm 1.03$      & $9.0 \times 10^{-6}$  \\
Inactive galaxies, $\alpha$  fixed      & 58.0$^{+99.8}_{-36.7}$      & $-1.50$              & $9.1 \times 10^{-6}$  \\
Active galaxies, $\alpha$  free         & 27.2$^{+1269.8}_{-26.6}$    & $-0.68\pm 1.49$      & $2.0 \times 10^{-4}$  \\
Active galaxies, $\alpha$  fixed        & 202.5$^{+521.9}_{-145.9}$   & $-1.50$              & $8.5 \times 10^{-4}$  \\
\enddata

\tablenotetext{a} {90\% confidence errors ($\Delta \chi^{2} = 2.71)$.}

\end{deluxetable}

\clearpage


\newpage

\begin{figure}
\epsscale{0.85}
\plotone{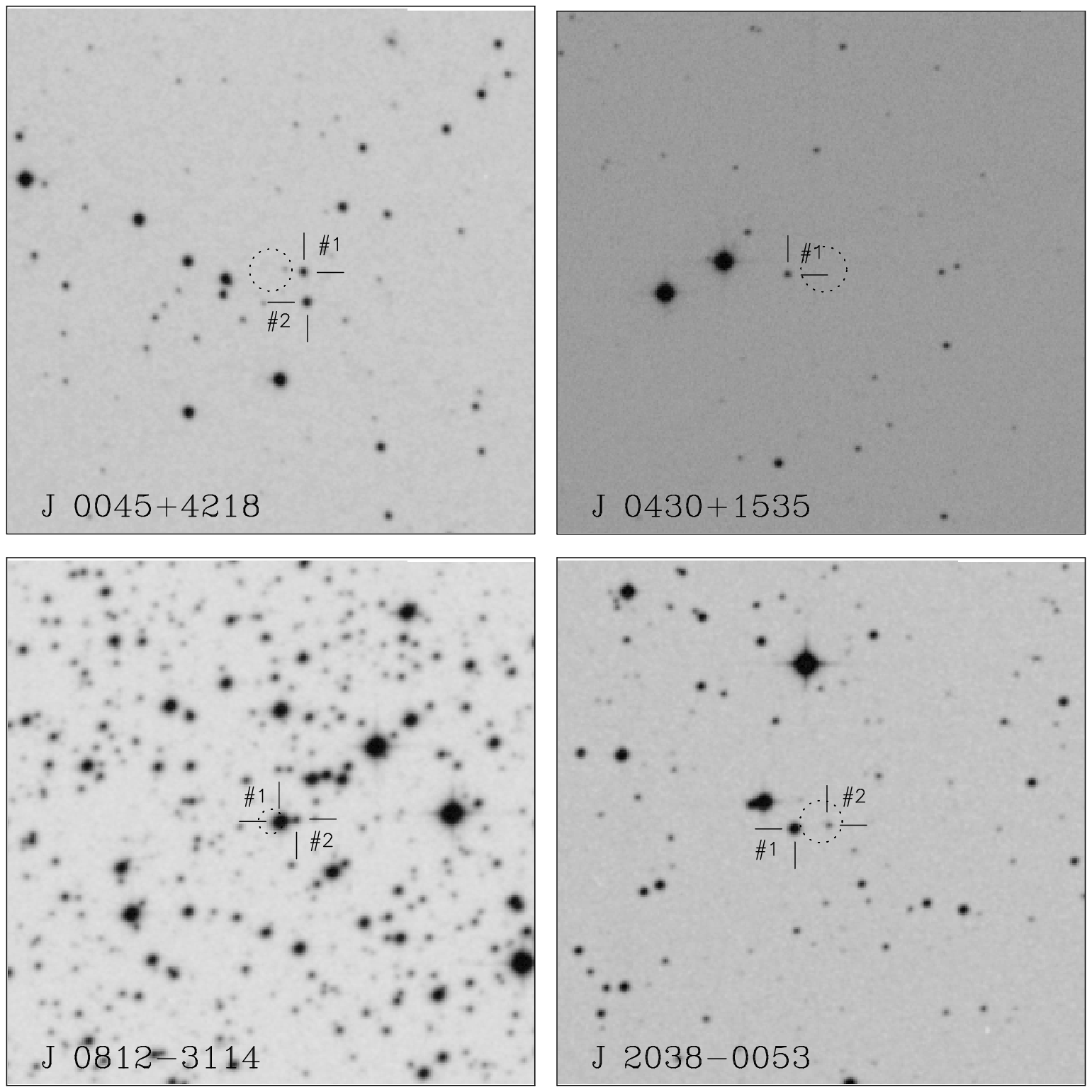}
\begin{flushleft}
Fig. A-1.--- Images of the fields of four flaring X-ray sources found
near the Galactic plane. These images were drawn from the 2nd
generation Palomar Digitized Sky Survey red plates. The size of each
field is $5^{\prime}$, and the coordinates of the center of the field
are given in Table~A-1. The X-ray source error circle is drawn as a
dotted circle, and the objects observed spectroscopically are
identified. The spectra of these objects are shown in Figure~A-2.
\end{flushleft}
\end{figure}


\newpage

\begin{figure}
\epsscale{0.85}
\plotone{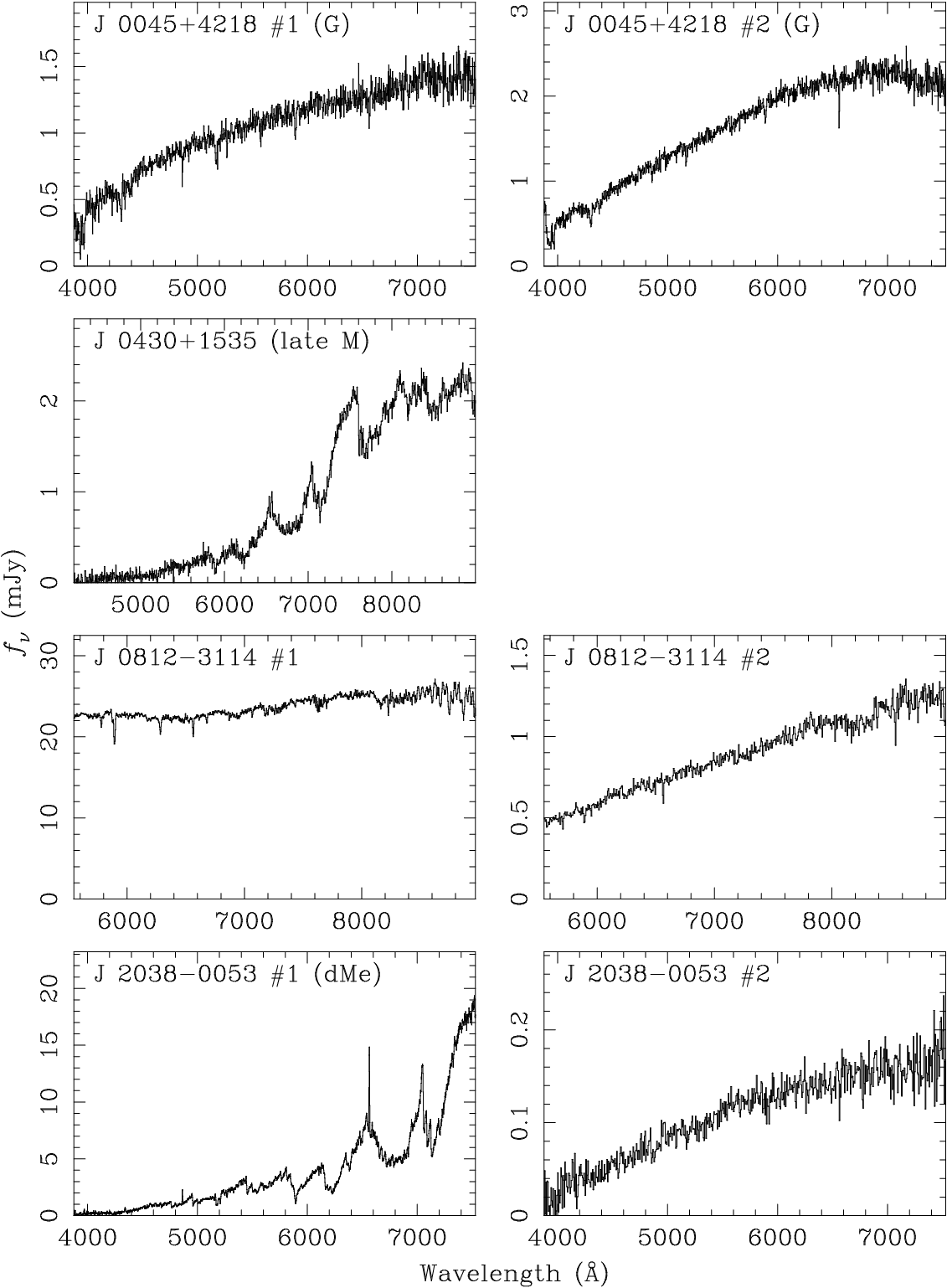}
\begin{flushleft}
Fig. A-2.-- Montage of spectra of optical objects in the immediate
vicinity of of flaring X-ray sources near the Galactic plane. These
objects are listed in Table~A-1 and identified in the images of
Figure~A-1.
\end{flushleft}
\end{figure}


\begin{deluxetable}{llllllllllllr}
\tablenum{A--1}
\tabletypesize{\tiny}
\tablewidth{0pt}
\rotate
\tablecaption{Log of \rosat\/ and Spectroscopic Observations}
\tablehead{
\colhead{}                                   &
\colhead{}                                   &  
\colhead{}                                   &
\colhead{\rosat\/}                           &    
\colhead{\rosat\/}                           &                            
\colhead{}                                   &
\colhead{Optical}                            &
\colhead{Telescope and}                      &
\colhead{Observation Date}                   &
\colhead{}                                  \\
\colhead{$\alpha_{2000}$}                    &              
\colhead{$\delta_{2000}$}                    &              
\colhead{Amp$_{\rm var}$\tablenotemark{a}}   &              
\colhead{Phase}                              &
\colhead{Date}                               &
\colhead{CR\tablenotemark{b}}                &              
\colhead{Source}                             &
\colhead{Instrument}                         &
\colhead{and time (UT)}                      &
\colhead{Time (s)}                          
}
\startdata

$00\ 45\ 28.7$  & $+42\ 18\ 50.0$  & 25  & RASS     &1990 July 12   &$0.059 \pm 0.021$            &\# 1 & KPNO 2.1m + Goldcam & 2000 Sep 24, 06:32:42 &  600 \\
                &                  &     & Pointed  &1993 Jan 31    &$\leq 0.00241$               &\#2  & KPNO 2.1m + Goldcam & 2000 Sep 24, 06:32:42 &  600 \\                                         

$04\ 30\ 38.3$  & $+15\ 35\ 19.3$  & 31  & Pointed  &1992 Dec 12    &$0.037 \pm 0.005$            &\#1  & HET + LRS           & 2000 Mar 15, 02:32:32 &780   \\       
                &                  &     & Pointed  &1992 Aug 10    &$0.0012 \pm 0.0008$          &     &                     &                       &      \\

$08\ 12\ 28.8$  & $-31\ 14\ 52.0$  & 78  & RASS     &1990 Oct 8     &$0.250 \pm 0.026$            &\#1  & CTIO 1.5m + CSPEC   & 2001 Jan 22, 08:28:32 & 1200 \\
                &                  &     & Pointed  &1992 Nov 20    &$\leq 0.0032$                &\#2  & CTIO 1.5m + CSPEC   & 2001 Jan 22, 08:28:32 & 1200 \\ 

$20\ 38\ 13.0$  & $-00\ 53\ 13.2$  & 56  & Pointed  &1991 May 11    &$0.870 \pm 0.022$            &\# 1 & KPNO 2.1m + Goldcam & 2000 Sep 24, 05:40:10 &  600 \\
                &                  &     & Pointed  &1992 Nov 16    &$0.016 \pm 0.001$            &\# 2 & KPNO 2.1m + Goldcam & 2000 Sep 24, 05:57:46 & 1800 \\

\enddata

\tablecomments{Units of right ascension are hours, minutes, and seconds, and units of declination are degrees, arcminutes, and arcseconds.} 
\tablenotetext{a} {Observed amplitude of count-rate variability in the 0.1--2.4 keV band.}
\tablenotetext{b} {Count rate in the $0.1$--$2.4$~keV energy band in cts~s$^{-1}$.}
\end{deluxetable}


\newpage

\begin{thebibliography}{}

\bibitem[]{}
 Allan, D. J., \& Vallance, R. J. 1995, ASTERIX X-Ray Data Processing
 System (Birmingham: Univ. Birmingham)

\bibitem[]{} 
Allan, D. J. 1995, {\sc Asterix} User Note 004: Source Searching and
Parameterization.  University of Birmingham, Birmingham

\bibitem[]{}
Angelini, L., Park, S., White, N. E., \& Giommi, P. 2000, AAS,
196.5310A

\bibitem[]{}
Armitage, P. J., Zurek, W. H., \& Davies, M. B. 1996, ApJ, 470, 237

\bibitem[]{}
Arnaud, K. A. 1996, in {ASP} Conf. Ser. 101, Astronomical Data
Analysis Software and Systems V, ed.  G.~Jacoby, \& J.~Barnes (San
Francisco: ASP), 17

\bibitem[]{}
Bade, N., Komossa, S., \& Dahlem, M. 1996, A\&A, 309, L35

\bibitem[]{}
Bade, N. et al. 1998, A\&AS, 127, 145

\bibitem[]{}
Boese, F. G. 2000, A\&AS, 141, 507

\bibitem[]{}
Brandt, W. N., Pounds, K. A., \& Fink, H. 1995, MNRAS, 273, L47

\bibitem[]{}
Brandt, W. N., Boller, Th., Fabian, A. C., \& Ruszkowski, M. 1999,
MNRAS, 303, L53

\bibitem[]{}
Burderi, L., King, A. R., \& Szuszkiewicz, E. 1998, ApJ, 509, 85

\bibitem[]{}
Cappellari, M., Renzini, A., Greggio, L., di Serego Alighieri, S.,
Buson, L. M., Burstein, D., \& Francesco, B. 1999, ApJ, 519, 117

\bibitem[]{}
Carrasco, L., Tovmassian, H. M., Stepanian, J. A., Chavushyan, V. H.,
Erastova, L. K., \& Vald\'{e}s, J. R.  1998, AJ, 115, 1717

\bibitem[]{}
de Jong, R. S.  1996, A\&A, 313, 45

\bibitem[]{}
de Vaucouleurs, G. The Observatory III, 1991, 122, No. 1102

\bibitem[]{}
Eracleous, M., Livio, M., Halpern, J. P., \& Storchi-Bergmann,
T. 1995, ApJ, 438, 610

\bibitem[]{}
Ferrarese, L. \& Merritt, D. 2000, ApJ, 539, L9

\bibitem[]{}
Gebhardt, K., et al. 2000, ApJ, 539, L13

\bibitem[]{}
Gehrels, N. 1986, AJ, 303, 336

\bibitem[]{}
Gehrels, N. 2000, X-ray and Gamma-Ray Instrumentation for Astronomy
XI, ed. K. Flanagan \& O. Siegnund (SPIE: Bellingham), 50

\bibitem[]{}
Greiner, J., Schwarz, R., Zharikov, S., \& Orio, M. 2000, A\&A, 362,
L25

\bibitem[]{}
Grupe, D., Beuermann, K. Mannheim, K., Bade, N., Thomas, H.-C., de
Martino, D. \& Schwope, A. 1995a, A\&A, 299, L5

\bibitem[]{}
Grupe, D., Beuermann, K., Mannheim, K., Thomas, H.-C., Fink, H. H., \&
de Martino, D. 1995b, A\&A, 300, L21

\bibitem[]{}
Grupe, D., Thomas, H.-C., \& Leighly, K. M. 1999, A\&A, 350, L31

\bibitem[]{}
Grupe, D. 2002, New Visions of the X-ray Universe in the XMM-Newton
and Chandra Era (ESA Press: Noordwijk), in press (astro-ph/0201048)


\bibitem[]{}
Guainazzi, M., Marshall, W., \& Parmar, A. N. 2001, MNRAS, 323, 75

\bibitem[]{}
Gurzadyan, V. G. \& Ozernoy, L. M. 1980, A\&A, 86, 315

\bibitem[]{}
Heiles, C. \& Cleary, M. N. 1979, Aust. J. Phys. Astrophys. Suppl.,
47, 1

\bibitem[]{}
Ho, L. C., Filippenko, A. V., Sargent, W. L. W., 1995, ApJS, 98, 477

\bibitem[]{}
Komossa, S. \& Dahlem, M. 2002, in MAXI Workshop on AGN Variability,
in press (astro-ph/0106422)

\bibitem[]{}
Komossa, S. \& Bade, N. 1999, A\&A, 343, 775

\bibitem[]{}
Komossa, S. \& Greiner, J. 1999, A\&A, 349, L45

\bibitem[]{}
Li, L. X., Narayan, R., \& Menou, K. 2002, ApJ, in press
(astro-ph/0203191)

\bibitem[]{}
Loeb, A. \& Ulmer, A. 1997, ApJ, 489, 573

\bibitem[]{}
Magorrian, J. \& Tremaine, S. 1999, MNRAS, 309, 447

\bibitem[]{}
Makishima, K. et al. 2000, ApJ, 535, 632

\bibitem[]{}
Micela, G., Sciortino, S., Kashyap, V., Harnden, F. R., Jr., \&
Rosner, R. 1996, ApJS, 102, 75

\bibitem[]{}
Mihara, T. et al. 2000, AdSpR, 25, 897

\bibitem[]{}
Mukai, K. 2000, PIMMS Version 3.0 Users' Guide. NASA/GSFC, Greenbelt

\bibitem[]{}
Packer, D. 1891, English Mechanic, 1389, 239

\bibitem[]{}
Parmar, A. N. 2001, The Observatory, 121, 1164

\bibitem[]{}
Peterson, B. M. 1997, An Introduction to Active Galactic Nuclei,
Cambridge University Press: Cambridge

\bibitem[]{}
Piro, L., Massaro, E., Perola, G. C. \& Molteni, D. 1988, ApJ, 325,
L25
 
\bibitem[]{}
Rees, M. J. 1990, Science, 247, 817

\bibitem[]{}
Renzini, A. et al. 1995, Nature, 378, 39

\bibitem[]{}
Schlegel, E. M. 1995, Rep. Prog. Phys. 58, 1375

\bibitem[]{}
Sembay, S. \& West, R. G. 1993, MNRAS, 262, 141

\bibitem[]{}
Siemiginowska, A., Czerny, B., \& Kostyunin, V. 1996, ApJ, 458, 491

\bibitem[]{}
Stark, A. A., Gammie, C. F., Wilson, R. W., Bally, J., Linke, R. A.,
Heiles, C., \& Hurwitz, M. 1992, ApJS, 79, 77

\bibitem[]{}
Storchi-Bergmann, T. Eracleous, M., Ruiz, M. T., Livio, M., Wilson,
A. S. \& Filippenko, A. V. 1997, ApJ, 489, 87

\bibitem[]{}
Syer, D., Clarke, C. J., \& Rees, M. J. 1991, MNRAS, 250, 505

\bibitem[]{}
Syer, D. \& Clarke, C. J. 1992, MNRAS, 255, 92

\bibitem[]{}
Veron-Cetty, M. P. \& Veron, P. 1998, A Catalogue of quasars and
active nuclei, 8th ed., Garching: European Southern Observatory (ESO),
ESO Scientific Report Series, vol 18

\bibitem[]{}
Voges, W. et al. 1999, A\&A, 349, 389
 
\end{thebibliography}
\end{document}